\documentclass[paper]{ieice}
\usepackage[pdftex]{graphicx,xcolor}
\usepackage[fleqn]{amsmath}
\usepackage{newtxtext}
\usepackage[varg]{newtxmath}
\usepackage{url}

\usepackage[ruled,vlined]{algorithm2e}
\usepackage{bbold}
\usepackage{tikz}
\usepackage{etoolbox}
\usepackage[caption=false,font=normalsize,labelfont=sf,textfont=sf]{subfig}

\usepackage{caption}

\field{}
\title{Safeguarding the Truth of High-Value Price Oracle Task: A Dynamically Adjusted Truth Discovery Method}

\authorlist{%
\authorentry[xianyouquan@stu.gxnu.edu.cn]{Youquan Xian}{n}{1,2}\MembershipNumber{}
\authorentry[liupeng@gxnu.edu.cn(Corresponding author)]{Peng Liu}{n}{1,2}\MembershipNumber{}
\authorentry[ldc@gxnu.edu.cn]{Dongcheng Li}{n}{1,2}\MembershipNumber{}
\authorentry[xyz@stu.gxnu.edu.cn]{Xueying Zeng}{n}{1,2}\MembershipNumber{}
\authorentry[wangp@gxnu.edu.cn]{Peng Wang}{n}{1,2}\MembershipNumber{}
}
\affiliate[1]{The author is with the College of Computer Science and Information Technology, Guangxi Normal University, Guangxi, China.
}

\affiliate[2]{The author is with the Key Lab of Education Blockchain and Intelligent Technology, Ministry of Education, Guangxi Normal University, Guangxi, China.
}

\received{2015}{1}{1}
\revised{2015}{1}{1}

\begin{document}
\maketitle
\begin{summary}
In recent years, the Decentralized Finance (DeFi) market has witnessed numerous attacks on the price oracle, leading to substantial economic losses. Despite the advent of truth discovery methods opening up new avenues for oracle development, it falls short in addressing high-value attacks on price oracle tasks. Consequently, this paper introduces a dynamically adjusted truth discovery method safeguarding the truth of high-value price oracle tasks. In the truth aggregation stage, we enhance future considerations to improve the precision of aggregated truth. During the credibility update phase, credibility is dynamically assessed based on the task's value and the Cumulative Potential Economic Contribution (CPEC) of information sources. Experimental results demonstrate a significant reduction in data deviation by 65.8\% and potential economic loss by 66.5\%, compared to the baseline scheme, in the presence of high-value attacks.
\end{summary}


\section{Introduction}\label{sec1}

In recent years, the DeFi market has faced significant challenges, with numerous attacks resulting in losses of hundreds of millions of dollars. The issue at the heart of these attacks is the manipulation of price oracles by attackers, resulting in transaction price differences and profiting \cite{oracle_news}. Oracles serve as a bridge between the blockchain and the outside world, making the reliability of their data a crucial cornerstone of DeFi development \cite{zhao2022toward}. It enables smart contracts to automate conditional execution operations based on external data by acquiring, verifying, and aggregating external data \cite{pasdar2023connect}.

The existing oracle schemes primarily concentrate on ensuring the credibility of nodes, data sources, and transmission processes. To guarantee node credibility, approaches like Augur \cite{peterson2015augur} and Astraea \cite{adler2018astraea} employ a voting game to determine answers, ensuring that honest node behavior is the optimal solution for their income. Extending this, Cai et al. \cite{cai2020truth,cai2022truthful} optimized voting weight and reward calculation methods to mitigate herd effects. Lin et al. \cite{lin2022novel} and Yu et al. \cite{yu2023lattice} devised oracle schemes based on the characteristic that threshold signatures necessitate at least threshold $t$ nodes to sign the answer for successful aggregation. Moreover, the characteristics of threshold signature broadcast ciphertext data can prevent the `Freeloading' problem\footnote{Node A copies the answer of node B without incurring any costs after cheating and seeing the feedback result of B. It undermines the diversity of data sources and discourages the oracle's prompt response.}.
Hence, threshold signatures find widespread application in commercial projects such as ChainLink \cite{chainlink} and DOS Network \cite{dos}. Similarly, Woo et al. \cite{woo2020distributed} and Liu et al. \cite{liu2022extending} employed Trusted Execution Environments (TEE) as an alternative to cryptography-based privacy protection technology, ensuring the correct execution of the logical code on node devices. Leveraging the reputation mechanism, Taghavi et al. \cite{taghavi2023reinforcement} and Xian et al. \cite{xian2023distributed} evaluated, selected, or rewarded high-quality oracle nodes, presenting an effective approach to enhance node credibility.
To guarantee the credibility of data sources, Lv et al. \cite{lv2021blockchain} and Almi'Ani et al. \cite{almi2023graph} employed a reputation mechanism and a weighted graph to assess the credibility of data sources, recommending reliable ones for tasks. However, this scheme may prove ineffective when both the data source and the node are untrusted. Solutions like DECO \cite{zhang2020deco} and SDFS \cite{he2019sdfs} ensure the security of the data transmission process through TLS and introduce a third party to validate that the data originates from the specified web data source, aiming to address the issue. Nevertheless, this approach contradicts the original goal of decentralization.

The oracle problem emerges when dealing with data sourced from untrusted sources and oracle nodes, complicating the determination of trustworthiness or factual representation. Fortunately, Truth Discovery (TD) serves as a technique that consolidates noisy information about the same set of objects or events, gathered from diverse sources, even in the presence of inconsistent or conflicting data, to ascertain the truth \cite{li2016survey,xiao2016towards,vadavalli2023novel}. Notably, Xiao et al. \cite{xiao2023decentralized} recently introduced the Decentralized Truth Discovery Oracle (DTDO), offering a solution for determining the truth of data in scenarios where both the oracle node and data source lack trustworthiness. It marks a significant stride in the evolution of oracle systems.

However, we observe that the aforementioned schemes often overlook the open and transparent nature of the blockchain and the emerging security challenges presented by varying transaction values in DeFi. Specifically, in payment transaction contracts, different profit scenarios entail different possibilities for profit-seeking users to maintain honesty\footnote{Potentially malicious users may not expose themselves in a \$1 task and choose to engage in malicious activities. However, in a \$1 million transaction, even long-time honest users may temporarily defect to secure a substantial profit.}.
The transparency of the blockchain makes transaction values publicly accessible, reducing the threshold for the aforementioned attacks. In this paper, we define the honesty of oracle users in long-term small transactions and the behavior of uploading abnormal data for high-value transactions as `high-value attacks', which reflects the actual idea of profit-seeking users in reality.
Traditional TD schemes, like DTDO \cite{xiao2023decentralized}, do not consider the occurrence of sudden attacks in the truth aggregation phase. Furthermore, its credibility update process treats each task equally and lacks effective memory of malicious behavior, which can potentially lead to substantial economic losses. Malicious nodes can swiftly restore their trustworthiness through engagement in numerous small-value transactional tasks.

To tackle the aforementioned challenges, we introduce a dynamically adjusted truth discovery method aimed at safeguarding the truth of high-value price oracle tasks. Initially, we enhance future considerations in the truth aggregation stage to improve the precision of truth estimation. Subsequently, in the credibility update phase, we dynamically adjust credibility based on factors such as task value and cumulative potential economic contribution of information source.

The contributions of this paper are summarized as follows:

\begin{itemize}
\item Aiming at the high-value attack of malicious users, we design a new truth aggregation method. Based on the idea of temporal difference algorithm, we add the consideration of future credibility in the truth aggregation stage to make the truth aggregation more accurate.
\item With the characteristics of high-value attacks in mind, we design a credibility update method that dynamically adjusts the credibility update weight based on the task value and the CPEC of the source. This adjustment aims to enhance the accuracy of source credibility evaluations.
\item Experiments show that our scheme can effectively reduce the data deviation by 65.8\% and the potential economic loss by 66.5\% compared to the baseline scheme in the presence of high-value attacks.
\end{itemize}

The rest of this paper is organized as follows. Section \ref{bg} introduces the oracle, truth discovery methods, and current problems. Section \ref{work} introduces the proposed scheme and provides an example and analysis. Section \ref{result} presents the experimental results and analysis. Finally, Section \ref{conclusion} concludes this paper and outlines ideas for future work.

\section{Preliminary and Existing Problems}
\label{bg}

\subsection{Oracle}

    \begin{figure}[h!]
        \centering
        \includegraphics[width=3in]{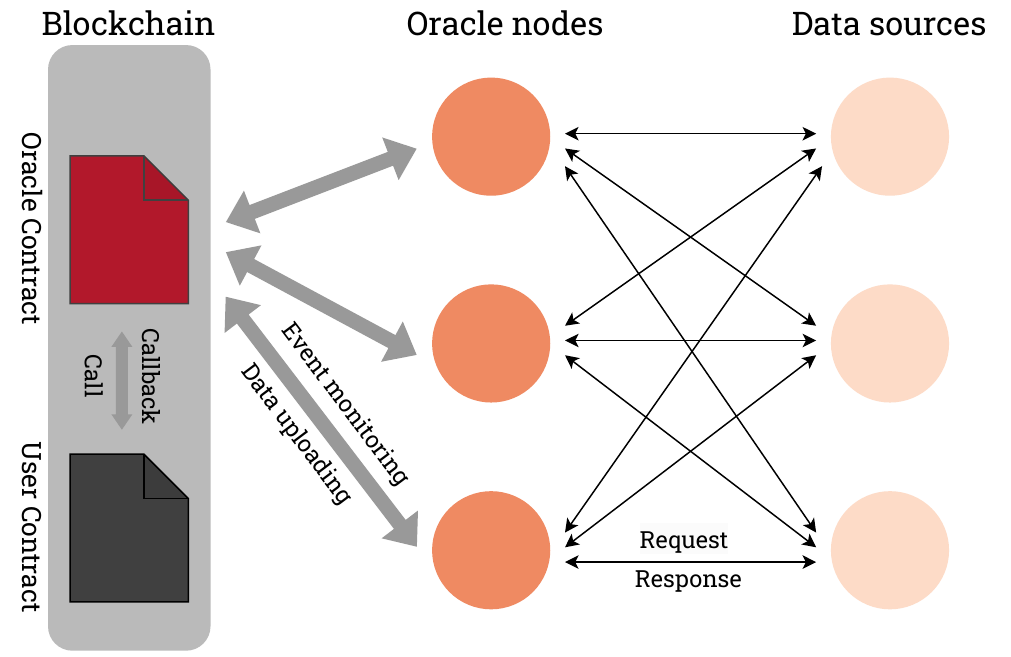}
        \caption{Overview of Oracle system.}
        \label{fig:Oracle}
    \end{figure}

Oracle is a trusted middleware that connects blockchain with real-world data, enabling smart contracts to perform automated conditional operations using external data. As shown in Fig. \ref{fig:Oracle}, it mainly consists of two parts: the network consisting of the off-chain oracle node (`node' hereafter) and the oracle contract. The user contract generates the corresponding request event on the blockchain by invoking the oracle contract interface, utilizing a triple $(I, S, F)$ to represent (request ID, data source set, and callback function).

The oracle network consists of $N$ nodes whose task is to provide data for blockchain applications. Each node $O_i$ ($i \in \{1,2,...,N\}$) continuously monitors the request event from the oracle contract and retrieves the data set $X_{i}^{I}( X_{i}^{I} \in \{ x_{i,j}^I \},j \in S)$ from $S$ according to the content of the event. According to the aggregation strategy, the final oracle contract is aggregated from the data set $X_{i}^{I}$ of $N$ nodes to obtain the final result $\hat{x}^{I}$. Finally, the oracle contract calls the callback function $F$ and returns the result to the user contract of the calling interface.

\subsection{Truth Discovery}
Truth Discovery (TD) is an algorithm for dealing with uncertain information, usually used to deal with inconsistencies, errors, or conflicting information in a data set. The goal is to find the results that are closest to the real situation by integrating information from several different sources. In the process of truth discovery, the estimation of source credibility and the step of truth aggregation are closely linked by the following principles: the source that provides more real information is given higher credibility, and the information supported by the reliable source is taken as the truth \cite{li2016survey}.

Given the source set $S$ and the obtained data object set $\{ x_{s} \}, s \in S$. TD aims to jointly estimate the truth $\hat{x}$ and the source credibility $\{r_s\}, s \in S$ behind the data object $x_{s}$, and to make the estimated result $\hat{x}$ as close as possible to the real truth $\check{x}$. These formulas usually take the form of a joint optimisation problem, the solution of which is similar to an iterative process in which truth aggregation (Eq. (\ref{eq1})) and source credibility estimation (Eq. (\ref{eq2})) are alternately performed until a certain convergence criterion is satisfied.

\begin{equation}
    \label{eq1}
    \hat{x}  = \frac{\sum_{s \in S}{\mathbb{1}_{s} \cdot  r_s \cdot x_{s}}}{\sum_{s \in S}{\mathbb{1}_{s} \cdot r_s}} 
\end{equation}
    
\begin{equation}
\label{eq2}
r_s = g \left (  \sum_{s \in S}{d(x_{s}, \hat{x} )} \right ) 
\end{equation}

Where $\mathbb{1}_{s}$ returns $1$ if the source $s$ provides the input (otherwise $0$), $d(\cdot)$ is the distance measure between the input $x_{s}$ and the current truth estimate $\hat{x}$, and $g(\cdot)$ is a monotone decreasing function. Different schemes have different choices for $d(\cdot)$ and $g(\cdot)$. In the iterative estimation process, the TD algorithm can evaluate the consistency of the output object $x_{s}$ of the source $s$ and assign high credibility to the source with high consistency. Note that both the data source and the oracle node can become the source $s$ in the oracle.

\subsection{Existing Problems}
\label{questions}

The basic assumption of the traditional TD algorithm is that the source that provides more real information is given greater credibility, and the information supported by the reliable source is taken as the truth. However, in applications with complex interests such as DeFi, the ordinary TD algorithm cannot cope with the attack of potentially malicious nodes on high-value tasks.

\subsubsection{\textbf{Question 1}}\label{q1} TD evaluates the credibility of the source according to its historical behaviour in the credibility evaluation phase, and unconditionally trusts the data of the highly credible source in the truth aggregation phase. However, in real-world scenarios, historical information does not fully reflect the current credibility of the source. For example, in the high-value attacks, even users who remain honest in historical tasks may decide to defect due to the huge benefits of the current task. The traditional TD scheme only considers the credibility of the node's history in the aggregation phase and cannot cope with this random and sudden attack.

\subsubsection{\textbf{Question 2}}\label{q2}
At the stage of evaluating the credibility of the source, traditional TD only considers data quality, but does not consider other important information about the task, including the task value, the potential economic loss caused by the source, and the lack of effective memory for malicious behaviour. 
As a result, traditional TD cannot defend against high-value attacks, because the malicious sources can be honest in many small-value transaction tasks, accumulate credibility, and suddenly behave badly in high-value transactions. It will cause large losses to the entire system, but the malicious sources will only temporarily lose some credibility, which can be quickly regained in subsequent tasks. Therefore, it is unreasonable to weight tasks of different values with the same credibility, and high-value tasks should be given more continuous attention. We show the actual performance of the above two problems in detail in Fig. \ref{fig:weight}.

\section{The proposed scheme}
\label{work}

In this section, we present in detail the dynamically adjusted truth discovery (DATD) method proposed in this paper.

\begin{figure*}[t]
    \centering
    \includegraphics[width=5in]{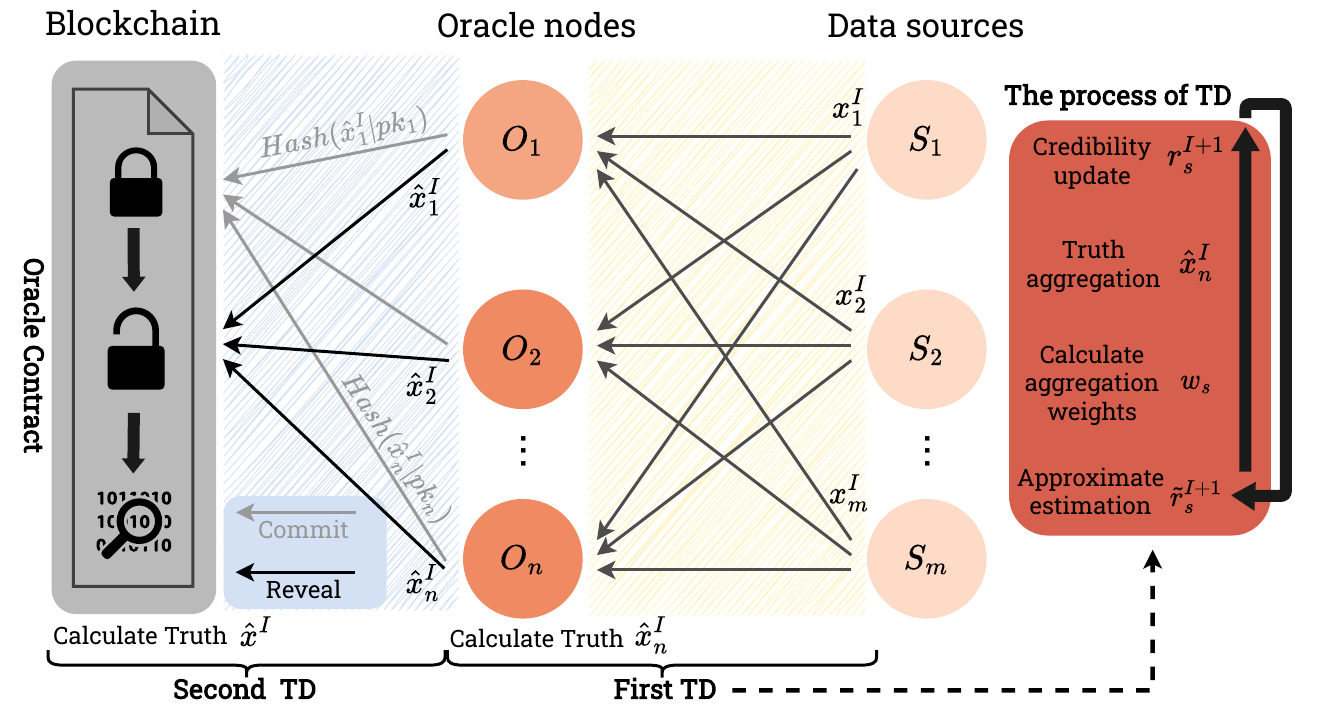}
    \caption{The process of two-stage TD in the proposed scheme.}
    \label{fig:overview}
\end{figure*}

\subsection{Overview}
\label{overview}
Specifically, considering the two key problems that exist at present, we design a new truth discovery method to protect the truth of the high-value price oracle. Compared with the traditional TD algorithm, our method is mainly optimised from the two stages of truth aggregation and credibility evaluation to improve the resistance of the TD algorithm to high-value attacks, which are introduced in detail in \$\ref{truth_aggregation} and \$\ref{credibility_degree_estimation}.

We divide an oracle task into the following four main stages, Fig. \ref{fig:overview} shows the process of the two most critical TD in the proposed scheme.

\begin{enumerate}
    \item \textbf{Task publishing}: A DeFi contract that needs to get the price data off-chain for trading will call the Oracle contract interface and generate a request event $(I,S,F,M_I)$ on the blockchain. Unlike normal oracle tasks, we add $M_I$ to indicate the value of the transaction. After the oracle node listens to the event, it requests data from the corresponding data source $s, s \in S$ according to the content of the event.
    \item \textbf{First TD}: The main purpose of the first TD is that the node $O_i, i\in O$ uses the TD algorithm to find its truth $\hat{x}_{i}^{I}$ from the data set $\{ x_{1}^{I},...,x_{m}^{I} \}$ from $S$.
    \item \textbf{Second TD}: The purpose of the second TD is similar to the first. The TD algorithm is used to find the truth $\hat{x}^{I}$ from the data set $\{ \hat{x}_{1}^{I},...,\hat{x}_{N}^{I} \}$ from $O_i$. Unlike DTDO, we use the commit-reveal protocol to upload data in the second TD to avoid freeloading problems. The commit stage first uploads a summary of the data $Hash(\hat{x}_{i}^{I} | pk_i)$, where $pk_i$ is the public key of node $O_i$. In the reveal phase, $(\hat{x}_{i}^{I},pk_i)$ is uploaded for matching verification.
    
    \item \textbf{Data callback}: The oracle contract calls the callback function $F$, and returns the final truth of the aggregation $\hat{x}^{I}$ to the user contract that invokes the interface.
\end{enumerate}

\textbf{System Postulation:} First, we assume that most data sources and oracle nodes are honest. Specifically, at any time in the system, there may be $f_s$ malicious data sources and $f_o$ malicious oracle nodes, $f_s < \frac{1}{2}$ and $f_o < \frac{1}{3}$, malicious data sources and nodes may have Byzantine failures, transmissions or manipulations. Then, we also assume that there are high-value attacks in real oracle environments, especially in applications like the proposed DeFi. That is, a smart malicious user will choose to be honest in low-value transactions and suddenly defect in high-value transactions, seeking the maximum value for his interests. However, the amount of his defection is related to his expected return and his real-time credit limit. This paper does not discuss it in detail and only assumes that the attack exists.

\subsection{Truth Aggregation}
\label{truth_aggregation}

\begin{algorithm}[t!]
    \caption{Approximate credibility estimation (Estimate)}
    \label{algorithm:estimate}
    \LinesNumbered
    \KwIn {The aggregation weight set of $S$, the data set obtained from $S$, and task value in round $I$: $\{w_s^{I}\}$, $\{x_s^I\}$, $M^I$}
    \KwOut {The set of credibility of approximate estimates: $\{\tilde{r}_s^{I+1}\}$}
    \BlankLine

    \tcp{truth aggregation}
    $\hat{x}^I_i  = \frac{\sum_{s \in S}{\mathbb{1}_{s}^{I} \cdot  w_{s} \cdot x_{s}^{I}}}{\sum_{s \in S}{\mathbb{1}_{s}^{I} \cdot w_{s}}}$\;   

    \tcp{credibility update}
    \For{$s \in S$}{
        $\sigma_s^{I} = \sqrt{(x_{s}^{I} - \hat{x}^I_i)^2}$ \;   
    }

    \For{$s \in S$}{
        $\sigma_s^{I} \gets \frac{\sigma_s^{I}}{\sum_{s \in S}\mathbb{1}_{s}^{I} \cdot \sigma_s^{I}}$  \;  
    }

    $\sigma^{I} = \sqrt{\frac{\sum_{s \in S}{ \mathbb{1}_{s}^{I} \cdot  (\sigma_s^{I})^2}}{\sum_{s \in S}{\mathbb{1}_{s}^{I}}}}$ \;

    \For{$s \in S$}{
         $\varrho_{s} = \frac{\sum_{i \in I}{\mathbb{1}_{s}^{i}}}{I} $ \;

         $d(x_{s}^{I},\hat{x}^{I}_i) = log_2(\sigma^{I} ) - log_2(\sigma_s^{I})$ \;
    }
    \For{$s \in S$}{
         $r_s^{I+1} = Sigmod(\varrho_{s} \frac{\sum_{i=1}^{I} M^{i} d(x_{s}^{I},\hat{x}^{I})}{M^{I}}) $ \;
    }
    
    \textbf{return} $\{r_s^{I+1}\}$;
\end{algorithm}

To solve question 1 (\$\ref{q1}), we use the idea of the temporal difference algorithm to increase the consideration of the future in the original truth aggregation stage, so that the estimation of the truth is more accurate. Because the source is credible in the past, it does not mean that the present and the future are credible. In the traditional TD method, if the source performs poorly in this task, its loss will only be reflected after this credibility update, but the truth aggregation stage of this round will also be influenced by its previous high credibility.

The basic idea of the temporal difference algorithm is to use the difference between the current estimated value and the actual value of the next step (temporal difference error) to update the estimated value at each step, so that the estimated value gradually approaches the actual value \cite{tesauro1995temporal}. Following this idea, we need to add future considerations to the calculation of the credibility $r_{s}$ of the source $s$, instead of directly using the past credibility $r_{s}^{I}$. Therefore, we use the existing $\{r_{s}^{I}\}$, the collected data $\{x_{s}^{I} \},s \in S$ to make a Monte Carlo approximation of the expectation of the credibility of the source $s$ in the next round and obtain $\tilde{r}_s^{I+1}$. As shown in the Algorithm. \ref{algorithm:estimate}, the $Estimate$ algorithm computation process includes the complete Eq.(\ref{predict}) - Eq.(\ref{update}).

\begin{equation}
\label{start}
\tilde{r}_s^{I+1} = Estimate(\{x_s^I\},\{r_s^{I}\}, M^I) \ \forall s \in S
\end{equation}

We design an aggregation weight $w_{s}$ for truth aggregation. The weight $w_{s}$ divides the credibility of the source $s$ into two parts: the historical credibility $r_{s}^{I}$ and the future credibility $\tilde{r}_s^{I+1}$. $\gamma$ is a discount factor that represents the discounted ratio of the credibility of the next moment to the present and also represents the importance that $w_{s}$ attaches to history and the future. The larger the $\gamma$, the more confidence in historical credibility, conversely, the more attention is paid to the future.

\begin{equation}
\label{new_weight}
w_{s} = \gamma \cdot r_s^I + (1-\gamma) \cdot \tilde{r}_s^{I+1}  \ \forall s \in S
\end{equation}

On this basis, the node $O_i$ predicts the current truth $\hat{x}^{I}_{i}$ based on the aggregation weight $\{w_{s}\},s \in S$ and the collected data $\{x_{s}^{I} \},s \in S$. Here $\mathbb{1}_{s}^{I}$ returns $1$ if source $s$ provides data in task $I$ (otherwise 0). Note that Eq.(\ref{start}) still uses $r_s^I$ as the aggregation weight.

\begin{equation}
 \label{predict}
\hat{x}^I_{i}  = \frac{\sum_{s \in S}{\mathbb{1}_{s}^{I} \cdot  w_{s} \cdot x_{s}^{I}}}{\sum_{s \in S}{\mathbb{1}_{s}^{I} \cdot w_{s}}} \ \forall i \in O
\end{equation}

Although the TD algorithm (the first TD and the second TD) is used twice in the proposed scheme. However, the first TD is that each node $O_i$ finds the truth $\hat{x}^I_{i}$ from the provided data $\{x_{s}^{I} \},s \in S$ according to the credibility record $\{r_s\},s \in S$ of the data source $s$. The second TD is the oracle contract based on the credibility record $\{r_i\},i \in O$ of the node $O_i$ from its input data $\{\hat{x}^I_{i}\}, i \in O$ to find the final truth $\hat{x}^I$. 

We distinguish the two-stage TD in the Table. \ref{tab:td2}. Both \$\ref{truth_aggregation} and \$\ref{credibility_degree_estimation} are based on the first TD, and the process of the second TD is similar, so we will not repeat it too much.

\begin{table}[h]
\centering
\caption{The difference between two-stage TD.}
\begin{tabular}{llll}
\hline
 & information source & aggregator &truth \\ \hline
First TD & datasources $S$ & node $O_i$ & $\hat{x}^I_{i}$ \\
Secend TD & oracle nodes $O$ & oracle contract &  $\hat{x}^I$ \\
\hline
\end{tabular}

\label{tab:td2}
\end{table}

\subsection{Credibility Update}
\label{credibility_degree_estimation}

To solve question 2 (\$\ref{q2}), we added the consideration of task value, participation rate, etc. to the process of credibility evaluation. In the traditional TD method, if the source is malicious, the credibility of the loss is only related to the degree of data deviation, and has nothing to do with the actual economic loss of the task. A malicious user can easily get the value of the task from the blockchain, and can cause a lot of economic loss by doing bad things to high-value tasks. But they can restore their credibility by behaving honestly on a large number of low-value tasks, which is obviously unreasonable.

First, we use the Euclidean distance between the data $x_{s}^{I}$ proposed by the source $s$ and the truth $\hat{x}^I_i$ estimated by $O_i$ to measure the degree of deviation $\sigma_s^{I}$. Then, considering the large difference in the degree of data deviation between different tasks, we need to standardise the degree of deviation of the source.

\begin{equation}
\sigma_s^{I} = \sqrt{(x_{s}^{I} - \hat{x}^I_i)^2} \ \ \forall s \in S
\end{equation}

\begin{equation}
\sigma_s^{I} \gets \frac{\sigma_s^{I}}{\sum_{s \in S}\mathbb{1}_{s}^{I} \cdot \sigma_s^{I}} 
\end{equation}

Then, to ensure that bad behavior is punished, we add the mean square error of this task as a measure. If the error $\sigma_s^{I}$ exceeds $\sigma^{I}$, credibility is deducted. $d(x_{s}^{I},\hat{x}^{I}_i)$ is a monotonically decreasing function that increases the credibility of the source for small deviations and decreases the credibility of the source for large deviations.

 \begin{equation}
\sigma^{I} = \sqrt{\frac{\sum_{s \in S}{ \mathbb{1}_{s}^{I} \cdot  (\sigma_s^{I})^2}}{\sum_{s \in S}{\mathbb{1}_{s}^{I}}}} 
\end{equation}

 \begin{equation}
d(x_{s}^{I},\hat{x}^{I}_i) = log_2(\sigma^{I} ) - log_2(\sigma_s^{I})
\end{equation}

Considering that tampering with exchange rate information creates arbitrage opportunities for attackers, especially in tasks such as exchange rate settlement, which is a common challenge for price oracles. We introduce $\mathcal{C}_s^I$ to capture the Potential Economic Contribution (PEC) induced by source $s$ in this oracle task. If the source is malicious, $\mathcal{C}_s^I$ is negative, indicating the potential economic loss caused by its actions. Conversely, it is positive, indicating the economic contribution made by its honest behavior. Moreover, $\mathcal{C}_s^I$ is proportionate to the task's value $M^{I}$ and the extent of data deviation $d(x_{s}^{I},\hat{x}^{I}_i)$. It implies that in high-value tasks, malicious behavior from the source results in a significant loss of credibility. Hence, even if malicious nodes act honestly in small-value transactions, it becomes challenging to offset the losses incurred by their behavior.

\begin{equation}
\mathcal{C}_s^I = d(x_{s}^{I},\hat{x}^{I}_i) \times M^I
\end{equation}

We also need to calculate the cumulative participation rate $\varrho_{s}$ of source $s$ as an indicator of its credibility. The higher the participation rate, the higher the credibility of its historical data.

\begin{equation}
\varrho_{s} = \frac{\sum_{i \in I}{\mathbb{1}_{s}^{i}}}{I} 
\end{equation}

Finally, we introduce a novel credibility evaluation method to update the credibility of source $s$. This method takes into account the cumulative participation rate $\varrho_{s}$ and the Cumulative Potential Economic Contribution (CPEC) $\vec{\mathcal{C}_s^I} = \sum_{i=1}^{I} \mathcal{C}_s^i$ of source $s$. The $Sigmoid$ function is employed to map real numbers to the interval $(0,1)$ and imparts a slower recovery speed to sources in the negative interval.

\begin{eqnarray}
\label{update}
 r_s^{I+1} = Sigmod(\varrho_{s} \cdot \frac{\vec{\mathcal{C}_s^I}}{M^{I}}  ) \ \forall s \in S
\end{eqnarray}

Truth discovery necessitates the iterative execution of Eq. (\ref{start}) and Eq. (\ref{update}) until a specified convergence criterion is met.

\subsection{Examples}
\label{example}

In this section, we provide an example to illustrate how the proposed TD algorithm works. We select $O_1$ in the first TD to obtain data from $\{s_1,..,s_5\}$ and utilize it as an example to find the truth.

\textbf{Parameter Setting:} The discount factor is set to $\gamma = 0.5$, with the current task value being $M^{I} = 8$. The CPEC is calculated as $\vec{\mathcal{C}_s^I} = 2.5$, and the cumulative participation rate is $\varrho_{s} = 1.0$. Refer to Table. \ref{tab:my-table} for additional details.

\begin{table}[h]
\centering
\caption{Some parameters involved in the process of  truth discovery from $S$ by $O_1$.}
\begin{tabular}{cccccl}
\hline
sources & $r_s^{I}$ & $x_{s}^{I}$ & $\tilde{r}_s^{I+1}$ & $d(x_{s}^{I},\hat{x}^{I}_1)$ & $r_s^{I+1}$ \\ \hline
$s_1$ & 0.8 & 1.0 & 0.617 & 0.265 & 0.641 \\ \hline
$s_2$ & 0.8 & 1.0 & 0.617 & 0.265 & 0.641 \\ \hline
$s_3$ & 0.8 & 1.0 & 0.617 & 0.265 & 0.641 \\ \hline
$s_4$ & 0.95 & 0.5 & 0.598 & -0.015 & 0.574 $\downarrow$ \\ \hline
$s_5$ & 0.95 & 0.4 & 0.480 & -0.464 & 0.462 $\downarrow$  \\ \hline
\end{tabular}
\label{tab:my-table}
\end{table}

As indicated in Table. \ref{tab:my-table}, we assume that the credibility $r_s^{I}$ of source $s$ in the initial task is $\{0.8,0.8,0.8,0.95,0.95\}$, with $s_4$ and $s_5$ being highly trusted sources. If this moment, the sources $s_4$ and $s_5$ suddenly defect and deliberately submit lower price data $x_s^{I}$ for profit, then according to Eq. (\ref{predict}), the truth $\hat{x}^I_1$ aggregated by $O_1$ using the conventional TD scheme is 0.757.

\begin{eqnarray}
\hat{x}^I_{1}  = \frac{(0.8\times1.0)+...+(0.95\times0.4)}{5} \\
 = 0.757
\end{eqnarray}

In our methodology, $\tilde{r}_s^{I+1}$ corresponds to the next round of credibility in the conventional TD algorithm. It aggregates the truth by considering the historical credibility $r_s^{I}$ as $w_s$ and updating it to derive the credibility for the subsequent round, as depicted in Eq. (\ref{start}). Although $\tilde{r}s^{I+1}$ can diminish the aggregation weight of malicious users in subsequent tasks, the aggregated data is still influenced in the current task. Therefore, we introduce a new aggregation weight $w_s$ in Eq. (\ref{new_weight}) to better capture the truth. To illustrate, let's consider the calculation of the aggregation weight $w_{5}$ for $s_5$:

\begin{eqnarray}
w_{5} = 0.5 \times 0.95 + (1-0.5) \times 0.48 \\
 = 0.715
\end{eqnarray}

\begin{eqnarray}
\hat{x}^I_{1}  = \frac{(0.708\times1.0)+...+(0.715\times0.4)}{5} \\
 = 0.774
\end{eqnarray}

At this juncture, the aggregated truth $\hat{x}^I_{1}$ by $O_1$ is 0.774 $\uparrow$. Computed using Eq. (6)-(9), the final data deviation $d(x_{s}^{I},\hat{x}^{I}_1)$ is $\{0.265,0.265,0.265,-0.015,-0.464\}$. Using $s_5$ as an example, the calculation for the update of the next round of credibility $r_5^{I+1}$ according to Eq. (\ref{update}) is as follows:

\begin{eqnarray}
r_5^{I+1} = \frac{1}{1 + e^-{ \frac{(1.0\times (2.5 + (-0.464\times8)))}{8} }} \\
= 0.462
\end{eqnarray}

\subsection{Analyses}
Under the assumption specified in \$\ref{overview}, we demonstrate the data credibility of the proposed scheme under high-value attacks and Freeloading problems.

\textbf{Proposition 1} (Ability to against high-value attacks): In the event of a sudden high-value attack during the system's operation, the proposed scheme can mitigate the impact of this unexpected attack on truth aggregation.

Proof Sketch: When a source $s \in S$ succumbs to the temptation of high-value tasks and chooses to behave maliciously, its returned data deviates from the truth. In the truth aggregation stage, the aggregation weight $w_s$ not only considers the historical credibility $r_s^{I}$ of $s$ but also estimates its credibility in the next round $\tilde{r}_s^{I+1}$ based on its current performance. Even if $s$ has maintained high credibility in the past, due to dishonest behavior in the current task, according to Eq. (\ref{start}), a very low $\tilde{r}_s^{I+1}$ will be assigned, resulting in a low final $w_s$. This effectively weakens the impact of $s$ on the current truth aggregation and subsequent tasks, as shown in \ref{example}. It is essential to note that the accuracy of the TD algorithm is based on the assumption that the majority of sources are honest \cite{xiao2023decentralized}. This ensures that in the truth aggregation stage of the first TD, there is always a probability of more than half of being close to the truth, and in the second TD, there is always a probability of more than $1-f_o$ of being close to the truth.

\textbf{Proposition 2} (Ability to against Freeloading problems): If there is a lazy oracle node in the network that submits data by directly copying uploaded data from others as its answer, the proposed scheme can effectively resist this undesirable behavior.

Proof Sketch: The second TD is divided into two stages. In the commit stage, all nodes only submit their data digests as proofs. In the reveal stage, the actual data and public keys are formally submitted to match the previous digests. Although nodes can observe the answers of other nodes in the reveal stage, the digest has already been submitted at this point. Therefore, this mechanism prevents nodes from directly copying plaintext data on the blockchain, eliminating the Freeloading problem.

\textbf{Complexity:} The communication overhead of the system primarily stems from the network transmission process of $O(N \times M)$, where $N$ and $M$ represent the number of nodes and data sources, respectively. The computational overhead for node $O_i$ is $O(M)$, primarily involving the aggregation and credibility update of the data from data sources $S$. The computational overhead of the oracle contract is $O(N)$, involving the aggregation and credibility update operations of the data by the nodes. Despite the addition of the phase submission operation, it does not significantly impact the overall computational complexity.

 \begin{figure*}[t!]
\centering
\subfloat[Data deviation when $\tau$=0.1]{\includegraphics[width=0.33\linewidth]{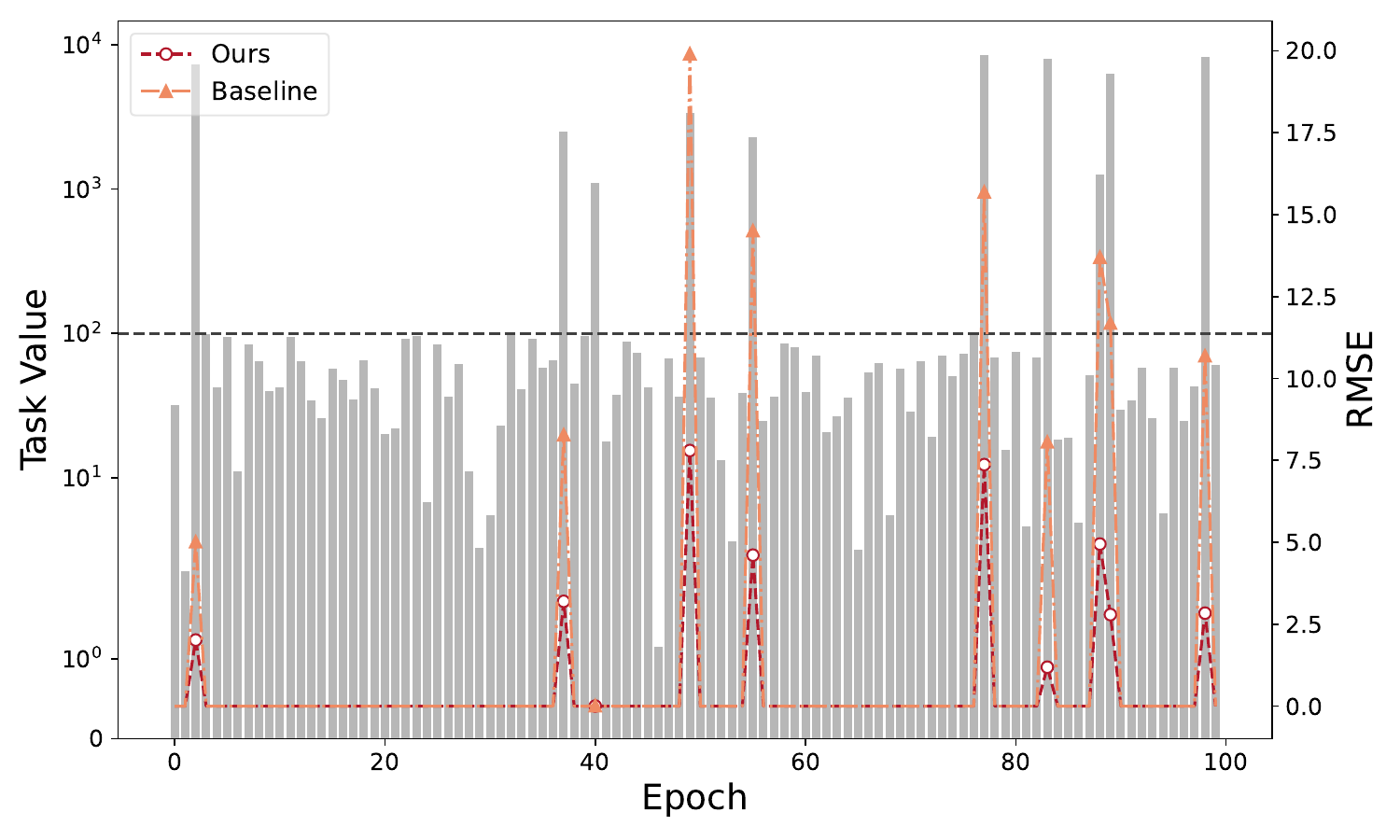}%
\label{fig:diff-0.9}}
\hfil
\subfloat[Data deviation when $\tau$=0.3]{\includegraphics[width=0.33\linewidth]{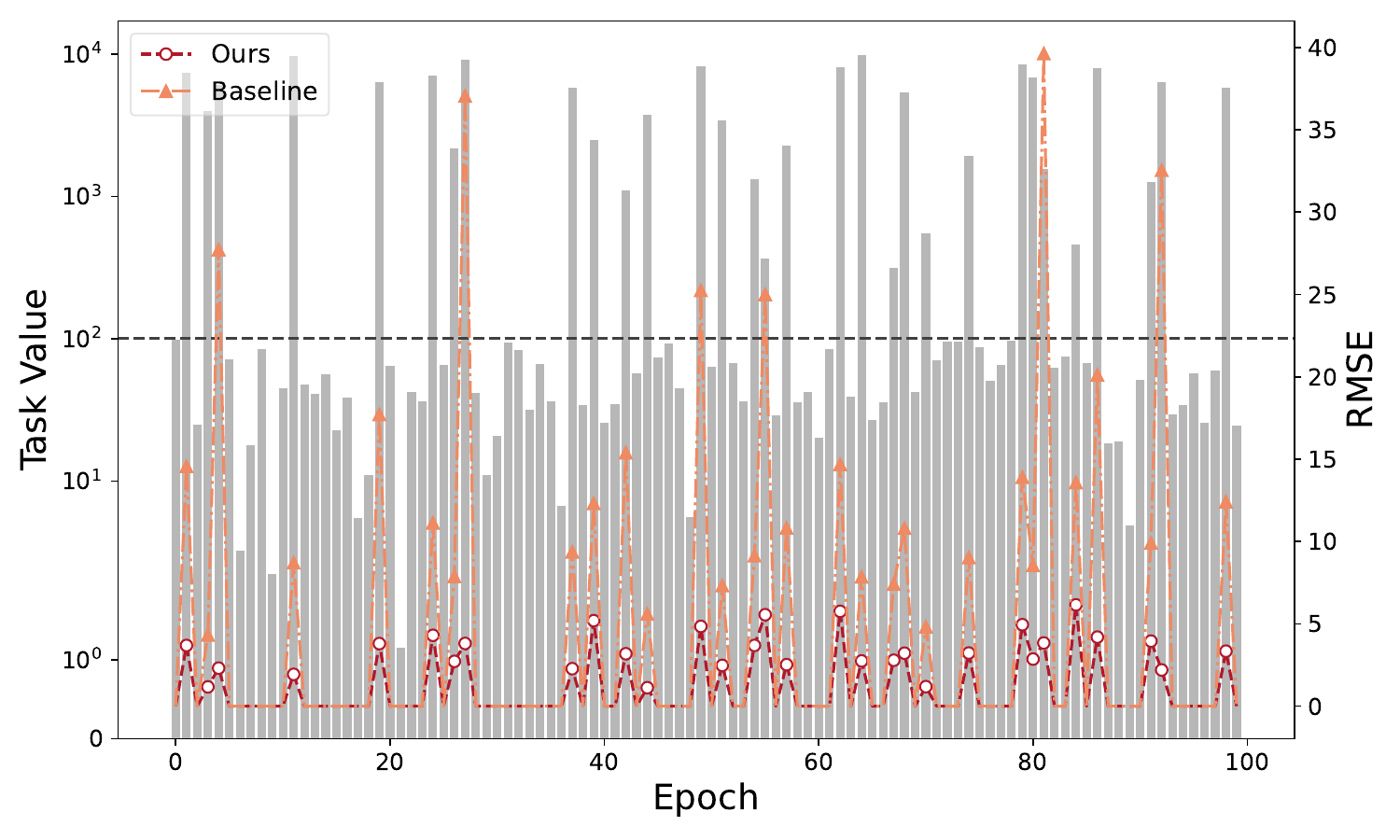}%
\label{fig:diff-0.7}}
\hfil
\subfloat[Potential economic losses when $\tau$=0.1]{\includegraphics[width=0.33\linewidth]{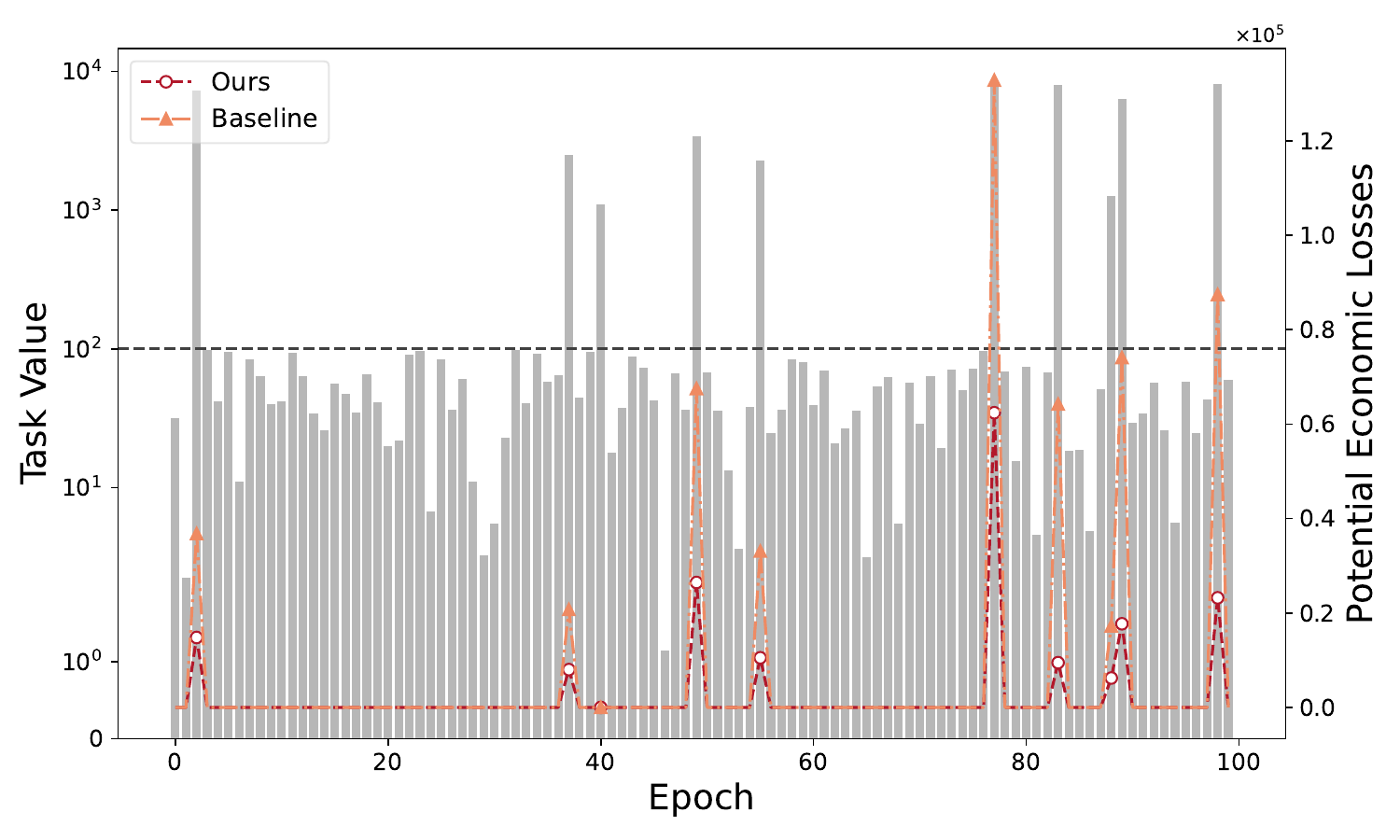}%
\label{fig:diff-value}}
\caption{Data deviation and economic loss under different conditions. (a)-(b) is the data deviation under high-value attacks of 10\% and 30\%; (c) is a possible economic loss under high-value attacks of 10\%.}
\label{fig:diff}
\end{figure*}

\section{Experiment and Result Analysis}
\label{result}

In this section, we employ Python to implement the prototype of the proposed method, encompassing two primary TD processes, as illustrated in Fig. \ref{fig:overview}.

\textbf{Experiment Setting:} In the established oracle network, there are 20 data sources and 20 oracle nodes. The proportions of malicious data sources and oracle nodes are denoted as $\alpha$ and $\beta$, respectively. The discount factor is set to $\gamma = 0.5$. The task of each oracle is to acquire random price data with values ranging from 0 to 100. High-value tasks have values between 100 and 10,000, while low-value tasks range from 1 to 100. Malicious sources maintain honesty in low-value tasks, while randomly adjusting price data in high-value tasks by (0-50\%) to seek profits. The baseline refers to the prototype implementation of DTDO \cite{xiao2023decentralized}. The experimental parameters are set in Table.\ref{tab:experimental-setup}.

\begin{table}[h]
\centering
\caption{Experimental parameter settings.}
\label{tab:experimental-setup}
\begin{tabular}{lll}
\hline
notation & meaning & value \\ \hline
$\gamma$ & discount factor & 0.5 \\ 
$\alpha$ & proportion of malicious data sources & 0.4 \\ 
$\beta$ & proportion of malicious oracle nodes & 0.3 \\ 
$\omega$ & range of malicious node data tampering & (0-50\%) \\
$\tau$ & proportion of high-value attacks & 0.1 \\
\hline
\end{tabular}
\end{table}

\subsection{Data Deviation}
Fig. \ref{fig:diff} illustrates the data deviation and potential economic loss across 100 tasks. Data deviation represents the extent of deviation between the TD estimated truth $\hat{x}^{I}$ and the actual truth $\check{x}^{I}$ of the task, calculated by the Root Mean Square Error (RMSE).

Firstly, in Fig. \ref{fig:diff-0.9}, the data deviation of each scheme is analyzed when facing 10\% high-value attacks in 100 tasks. The experiments reveal that the proposed scheme, on average, reduces data deviation by 65.8\% compared to the baseline during high-value attacks. Additionally, for each high-value attack (identified in the legend), the proposed scheme exhibits lower data deviation than the baseline.

Next, in Fig. \ref{fig:diff-0.7}, the data deviation of each scheme is presented under a 30\% high-value attack. Comparing with Fig. \ref{fig:diff-0.9}, it is observed that even in the scenario of frequent attacks, the truth value aggregated by the proposed scheme remains closer to the actual truth than the baseline.

Finally, Fig. \ref{fig:diff-value} analyzes the potential economic losses of each scheme when subjected to high-value attacks, represented by $\sqrt{(\check{x}^{I} - \hat{x}^{I})^2} \times M^I$. The results demonstrate that the proposed scheme reduces the potential economic loss by 66.5\% on average compared with the baseline.

\begin{figure}[h]
    \centering
    \includegraphics[width=3in]{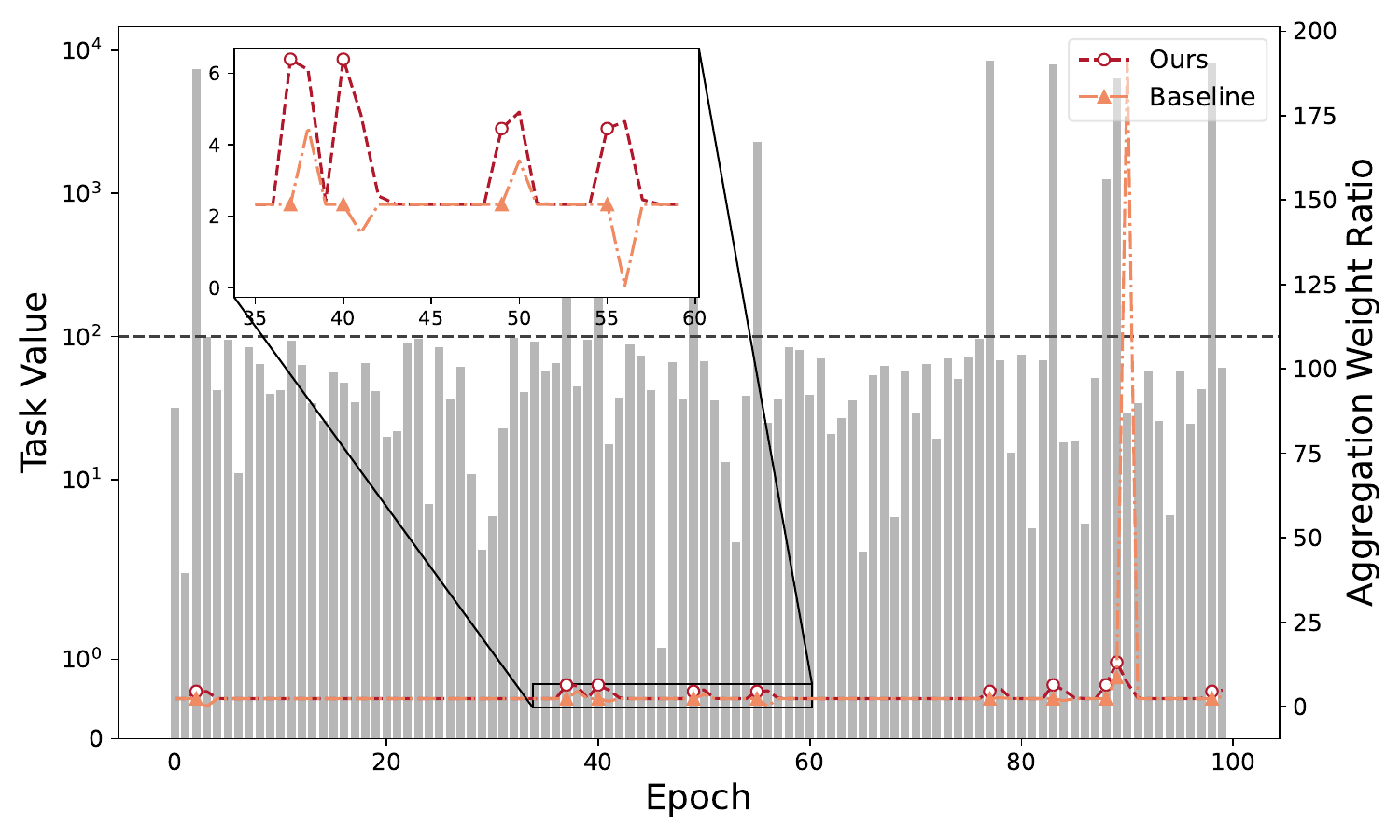}
    \caption{The aggregation weight for each truth aggregation (Second TD).}
    \label{fig:weight}
\end{figure}

\begin{figure}[h]
    \centering
    \includegraphics[width=3in]{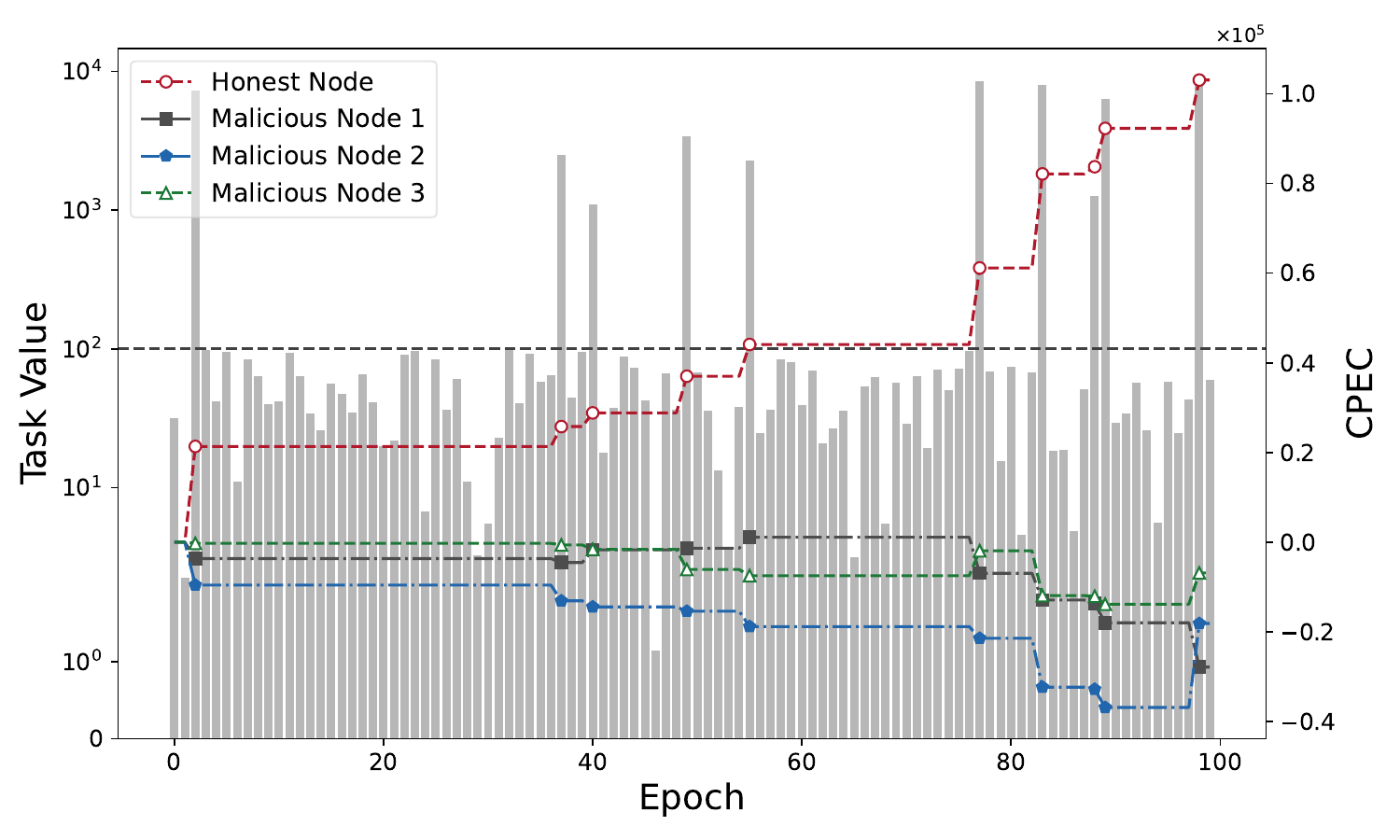}
    \caption{Changes in CPEC of nodes.}
    \label{fig:history}
\end{figure}

\begin{figure*}[t!]
\centering
\subfloat[Data deviation with different $\beta$]{\includegraphics[width=0.33\linewidth]{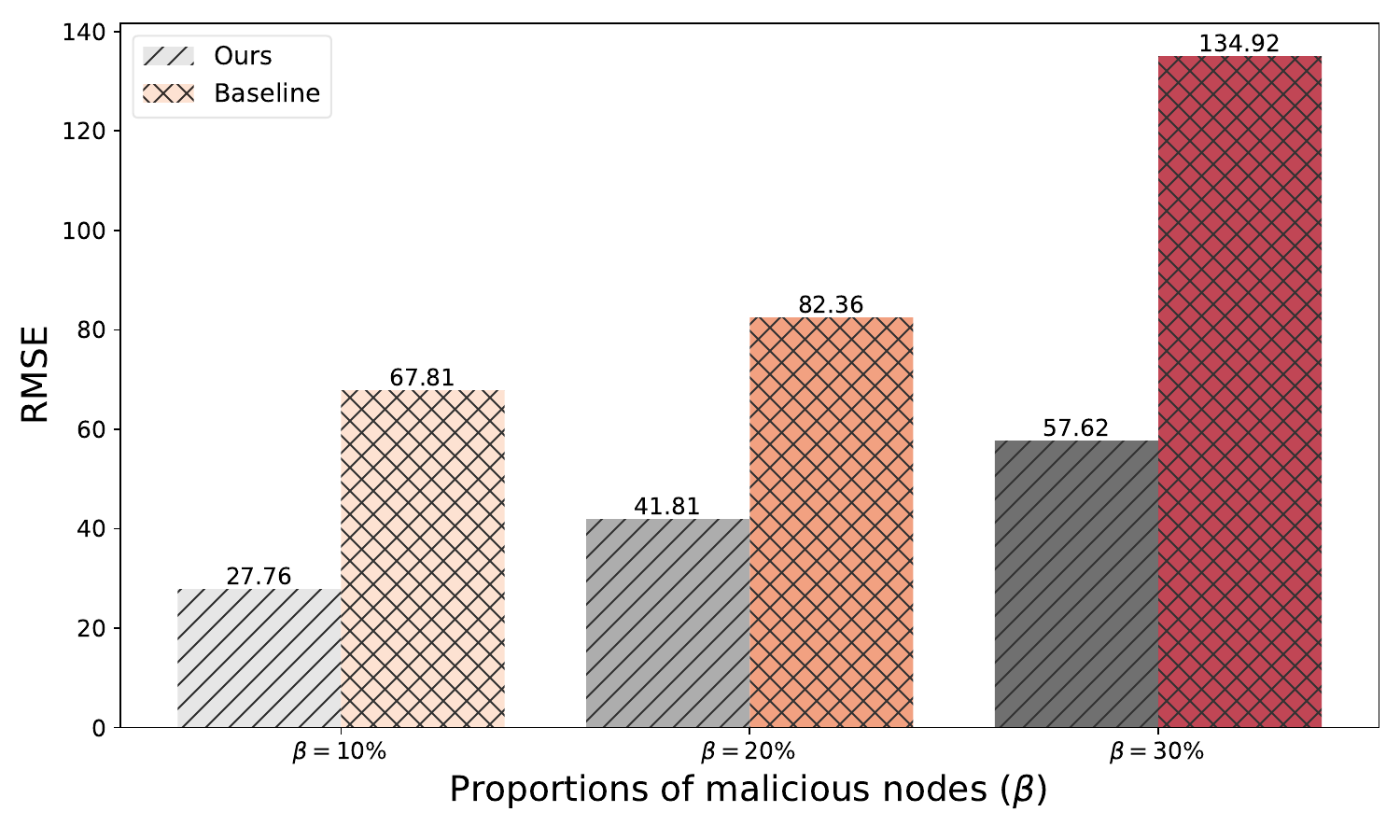}%
\label{fig:robustness-beta}}
\hfil
\subfloat[Data deviation with different $\omega $]{\includegraphics[width=0.33\linewidth]{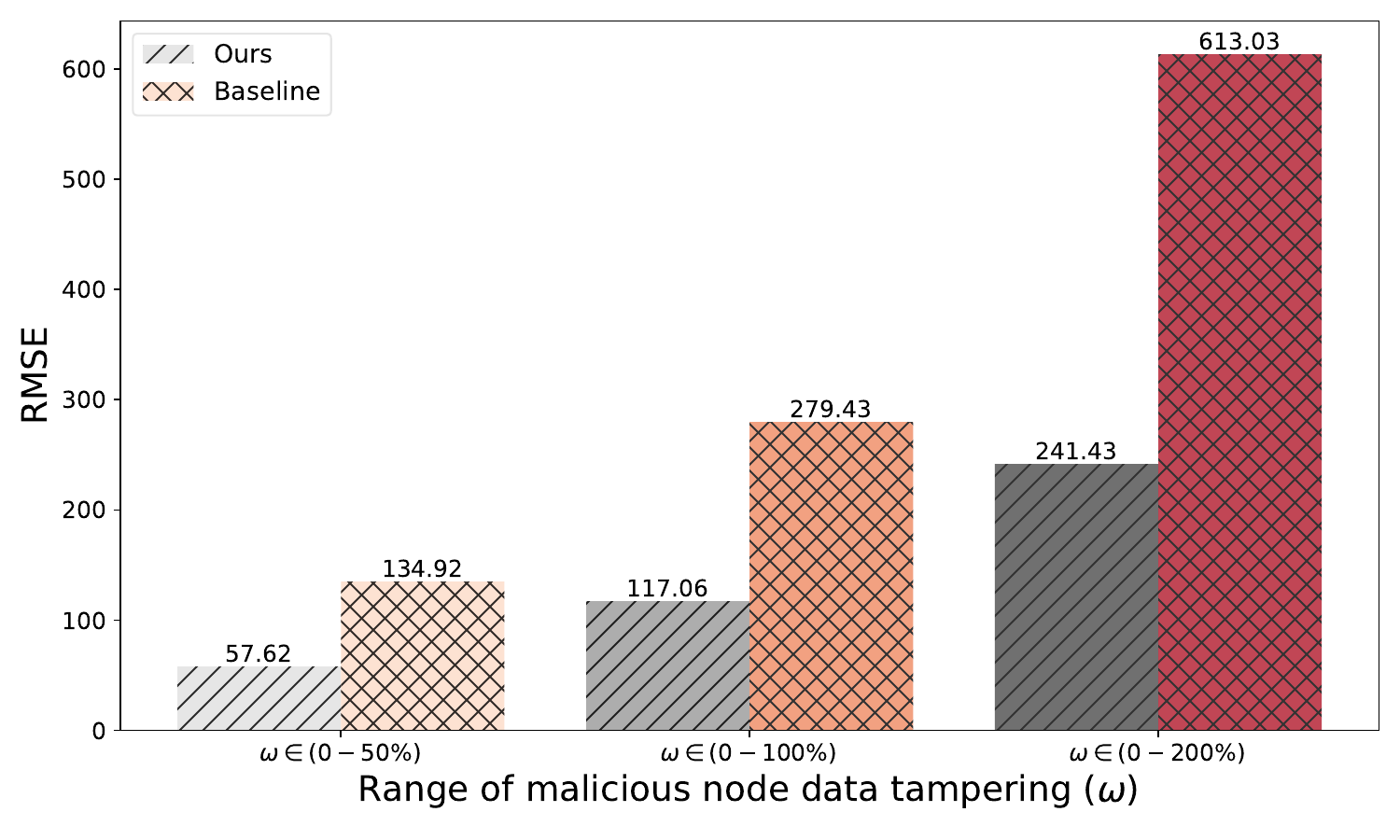}%
\label{fig:robustness-omega}}
\hfil
\subfloat[Data deviation in the first TD]{\includegraphics[width=0.33\linewidth]{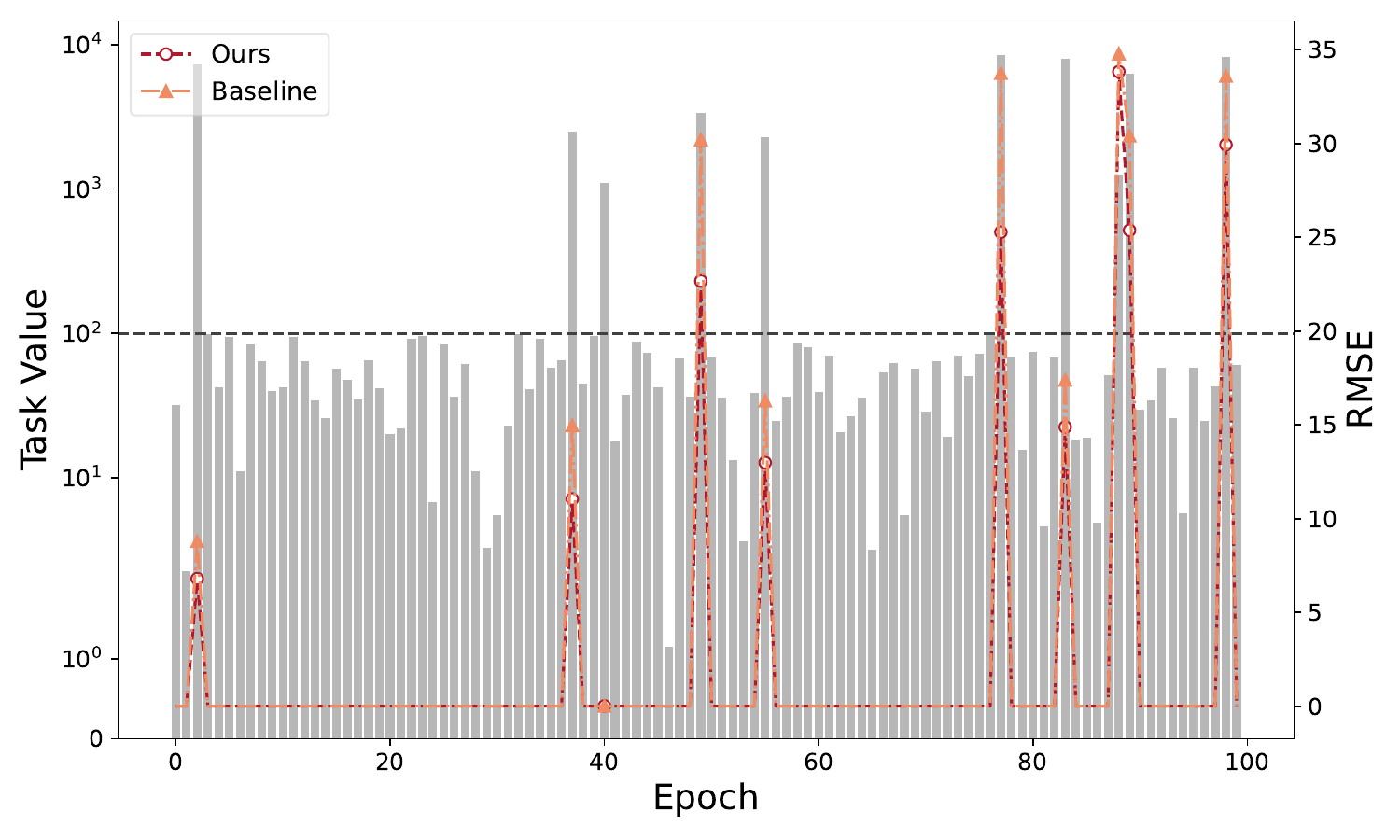}%
\label{fig:robustness-data-diff-0.9}}
\caption{The robustness of the proposed solution. Data deviation in (a) different proportion of malicious nodes $\beta $, (b) different range of malicious node data tampering $\omega$, and (c) the first TD.}
\label{fig:robustness}
\end{figure*}

\subsection{Aggregation Weight and Credibility} 

In order to explore why our scheme can protect the truth of high-value tasks, we analyze in detail the truth aggregation process of each round of tasks and the change of CPEC.

Fig. \ref{fig:weight} presents the aggregation weights of different schemes in each round of the truth aggregation stage along with their variations. The aggregation weight ratio in the graph indicates the ratio of the aggregation weight of an honest node to that of a malicious node. A higher ratio signifies a more accurate final aggregation result. Notably, in high-value tasks (as identified in the marker), the proposed scheme consistently maintains a higher weight ratio compared to the baseline. Furthermore, it is observed that the baseline consistently exhibits a slow reactive response to high-value attacks, potentially leading to misjudgments, aligning with our analysis in question 1 (\$\ref{q1}). Lastly, under sustained malicious activity, such as around the 90th round in the figure, the proportion of the baseline rises rapidly, indicating an increased identification of malicious nodes. However, after 1-2 rounds of honest behavior, the proportion rapidly decreases, illustrating a swift forgetting of historical malicious behavior, in line with our analysis in question 2 (\$\ref{q2}).

Fig. \ref{fig:history} illustrates the evolution of Cumulative Potential Economic Contribution (CPEC) $\vec{\mathcal{C}_s^I}$ for a randomly selected honest node and three malicious nodes over 100 tasks. As the number of iteration rounds increases, the discrimination of CPEC between malicious nodes and honest nodes also gradually intensifies. It implies that even if a malicious node exclusively engages in high-value tasks, its aggregation weight will gradually diminish due to the persistent memory of its historical behavior.

\subsection{Robustness}
To assess the robustness of the proposed scheme, we conducted tests on data deviation under various conditions, including different schemes, malicious nodes ratios $\beta$, malicious strategy (modification range) $\omega$, and the first TD process. The results are presented in Fig. \ref{fig:robustness}.

Fig. \ref{fig:robustness-beta} displays the total data deviation of different schemes when confronted with high-value attacks at varying proportions of malicious nodes $\beta=$ [10\%, 20\%, 30\%]. Notably, the proposed scheme consistently sustains a lower data deviation than the baseline and remains closer to the truth across different proportions of malicious nodes.

Xian et al. \cite{xian2023data} suggested that malicious nodes might seek benefits by adjusting the scope of data tampering. Therefore, we conducted an analysis of total data deviation when the range of malicious node data tampering, denoted as $\omega$, is (0-50\%), (0-100\%) and (0-200\%) in Fig. \ref{fig:robustness-omega}. The results reveal that even when malicious nodes employ different data tampering strategies, the proposed scheme consistently exhibits lower data deviation than the baseline.

To assess the effectiveness of the proposed TD scheme in the first TD phase, i.e., the truth discovery of the data source, we randomly selected a TD process of a node for analysis, exemplified in Fig. \ref{fig:robustness-data-diff-0.9}. Similar to Fig. \ref{fig:diff-0.9}, the proposed scheme demonstrates lower data deviation compared to the baseline and achieves more accurate truth discovery with smaller deviation in the first and second TD.

\begin{figure}[h]
    \centering
    \includegraphics[width=3in]{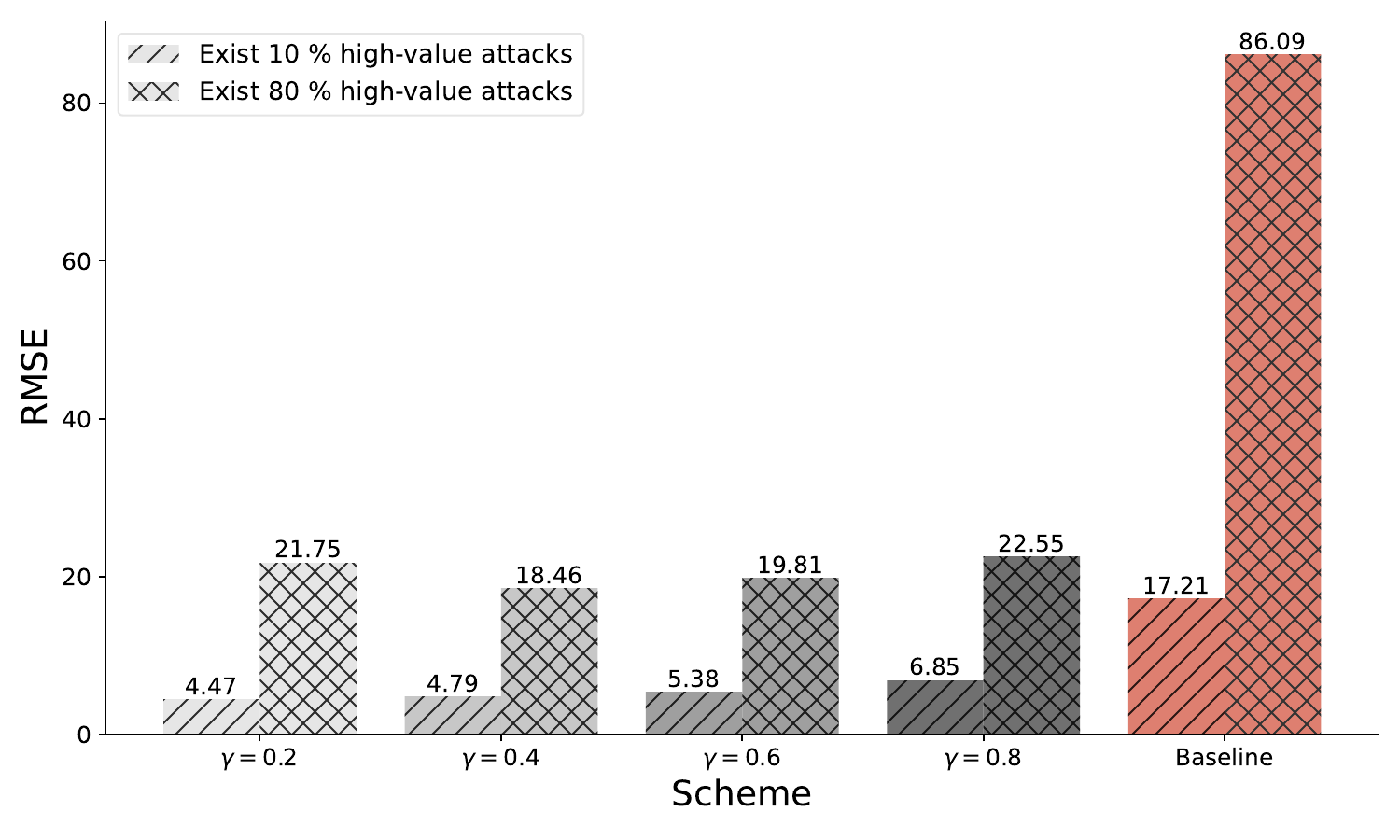}
    \caption{The influence of different $\gamma$ on TD.}
    \label{fig:gamma}
\end{figure}

Finally, to analyze data consistency with different values of $\gamma$ in the proposed scheme, we set $\gamma$ to [0.2, 0.4, 0.6, 0.8] and calculate the total data deviation for 10 tasks. Experimental results indicate that when 10\% of tasks are attacked by high-value attacks, data deviation increases with the rise of $\gamma$. This aligns with our assumption that increasing attention to the future enhances the node's ability to respond to sudden attacks. However, when attacks are more frequent, such as when there are high-value attacks in 80\% of tasks, proper consideration of historical credibility can help strengthen the identification of malicious nodes. Finally, no matter how the environment changes, the proposed scheme can always provide better data than baseline.

\section{Conclusion}
\label{conclusion}
This paper introduces a dynamically adjusted truth discovery method aimed at safeguarding the truth of high-value price oracle tasks. To enhance resistance against high-value attacks in the truth aggregation stage, we incorporate the concept of temporal difference algorithms, increasing consideration of future credibility during the aggregation stage to minimize truth aggregation deviation. Subsequently, in the credibility update stage, we dynamically adjust credibility based on factors such as task value and cumulative potential economic contribution of the sources, improving source evaluation accuracy. Experimental results demonstrate that in the presence of high-value attacks, the proposed scheme exhibits lower data deviation compared to the baseline scheme and reduces potential economic losses.

In the future, we plan to conduct a more in-depth exploration of the motivations and optimal strategies of malicious users, design more robust defense strategies against potential attacks, imporving the credibility of price oracles. Additionally, we aim to expand the truth discovery method to incorporate various types of data, including text, images, and other forms, improving its versatility and applicability.

\section*{Acknowledgements}
The research was supported in part by the Guangxi Science and Technology Major Project (No. AA22068070), the National Natural Science Foundation of China (Nos. 62166004,U21A20474), the Basic Ability Enhancement Program for Young and Middle-aged Teachers of Guangxi (No.2022KY0057, 2023KY0062), the Key Lab of Education Blockchain and Intelligent Technology, the Center for Applied Mathematics of Guangxi, the Guangxi "Bagui Scholar" Teams for Innovation and Research Project, the Guangxi Talent Highland Project of Big Data Intelligence and Application, the Guangxi Collaborative Center of Multisource Information Integration and Intelligent Processing.

\bibliographystyle{ieicetr}
\bibliography{myref}

\profile[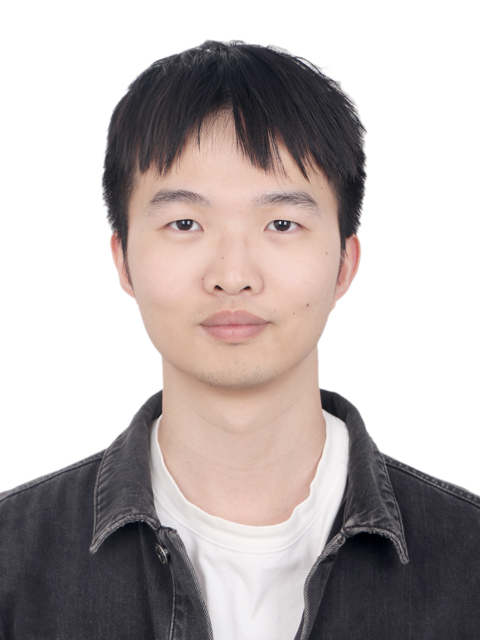]{Youquan Xian}{received the BEdegree from the BeiBu Gulf University, in 2021. He is currently working toward the master’s degree at Guangxi Normal University. His research interests include blockchain, edge computing, and federated learning.}
\label{profile}

\profile[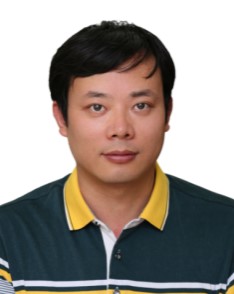]{Peng Liu}{received his Ph.D. degree in 2017 from Beihang University, China. He began his academic career as an assistant professor at Guangxi Normal University in 2007 and was promoted to full professor in 2022. His current research interests are focused on federated learning and blockchain.}
\label{profile}

\profile[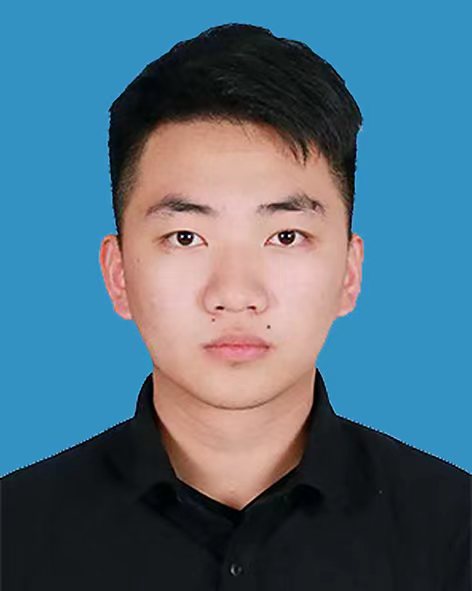]{Dongcheng Li}{received the master's degree in sofrware engineering from  Guangxi normal university. He  is currently working at the department of 
Computer Science and Information Technology of
Guangxi normal university, 
China. His main research interests include blockchain, data security and recommendation system.}
\label{profile}

\profile[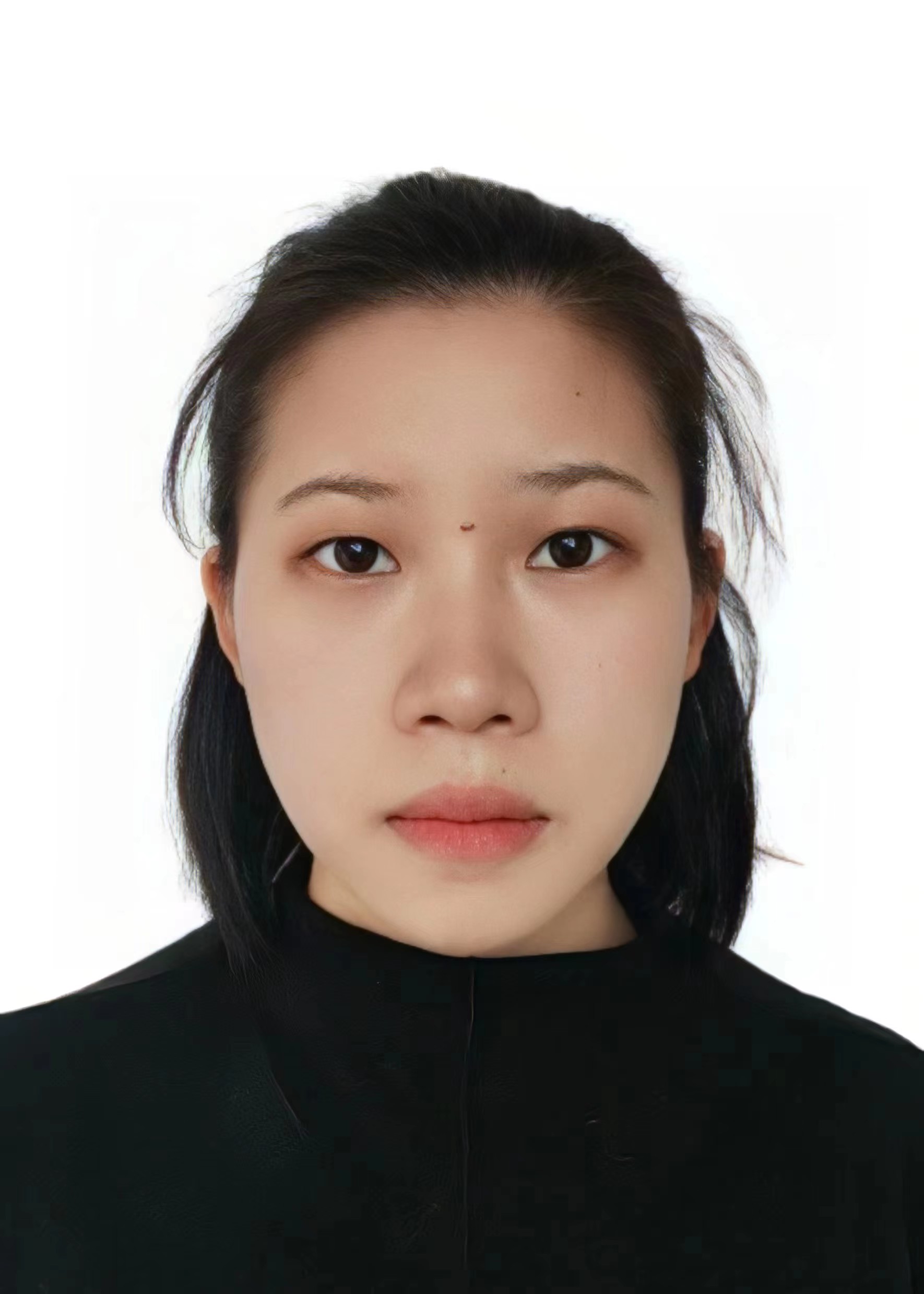]{Xueying Zeng}{received her bachelor's degree from Guangxi Science and Technology Normal University in 2022. She is currently pursuing her master's degree from Guangxi Normal University. Her research interests are blockchain, crowdsourcing.}
\label{profile}

\profile[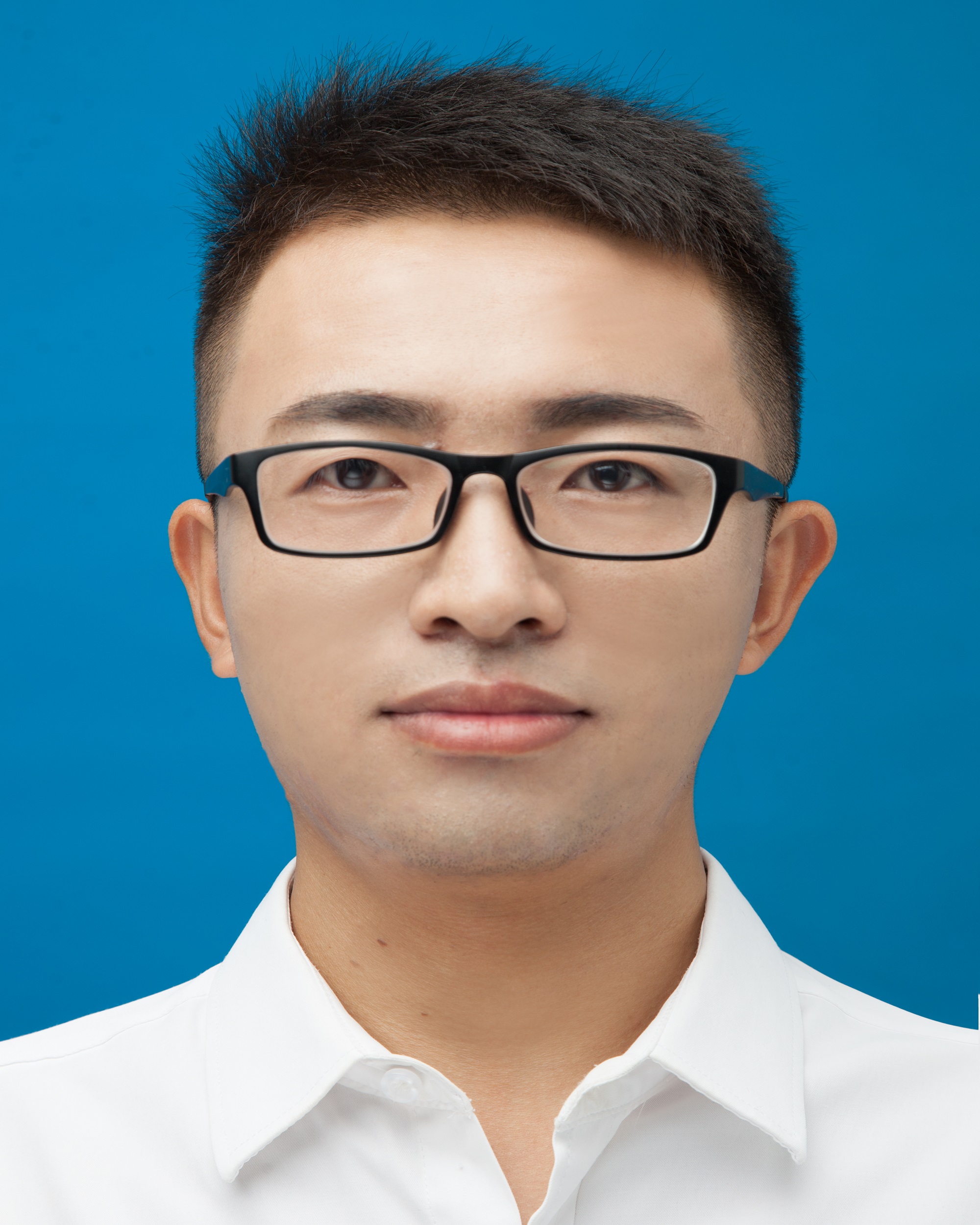]{Peng Wang}{received his master's degree from Guilin University of Technology in 2018. He is currently working toward a doctor's degree at Guangxi Normal University. His research interests include blockchain, data fusion, and data security.}
\label{profile}

\end{document}